# Elasticity on Ontology Matching
# of Folksodriven Structure Network

Massimiliano Dal Mas

*me @ maxdalmas.com*

**Abstract.** Nowadays folksonomy tags are used not just for personal organization, but for communication and sharing between people sharing their own local interests. In this paper is considered the new concept structure called "Folksodriven" to represent folksonomies. The Folksodriven Structure Network (FSN) was thought as folksonomy tags suggestions for the user on a dataset built on chosen websites – based on Natural Language Processing (NLP). Morphological changes, such as changes in folksonomy tags chose have direct impact on network connectivity (structural plasticity) of the folksonomy tags considered. The goal of this paper is on defining a base for a FSN plasticity theory to analyze. To perform such goal it is necessary a systematic mathematical analysis on deformation and fracture for the ontology matching on the FSN. The advantages of that approach could be used on a new interesting method to be employed by a knowledge management system.

**Keywords:** Semantic Web, Folksonomy, Ontology, Network, Elasticity, Plasticity, Natural Language Processing , Physics and Society, Quasicrystal

## 1 Introduction

Folksonomy is used as a system derived from user-generated electronic tags or keywords that annotate and describe online content. On the paper [1] was proposed the use of folksonomies and network theory to devise a new concept: a "Folksodriven Structure Network" to represent folksonomies.

The network structure of Folksodriven tags – *Folksodriven Structure Network* (*FSN*) – was thought as a "Folksksonomy tags suggestions" based on *Natural Language Processing (NLP)*, for the user on a dataset built on chosen websites. A *FSN* can be considered as a way to solve the Ontology Matching problem between Folksodriven tags – that are hard to categorize.

Analyzing *FSN* it was showed as it can facilitate serendipitous discovery of content among users. A *FSN* determines the flow of information in folksonomy tags network and determines its functional and computational properties [2].

A folksonomy activation of the tag goes along with a semantic increase in the web source classified by the user which induces morphological alterations of the folksonomy tag chosen on a network connection. Morphological changes, such as changes in folksonomy tags chose have direct impact on network connectivity (structural plasticity) of the sources considered (e.g. the website considered).

As a consequence of morphological changes, links between network tags may break and new links can be generated. Local structural changes at the single *FD tag* may entail alterations in global network topology of the *FSN*. Conversely, global topology can have impact on local chose on *FD tag* by the user on a new web source to correlate with a folksonomy tag.

These complex reciprocal interactions between structural changes and activity dynamics as well as local and global effects of structural plasticity necessitate theoretical modeling approaches to elucidate rapidly growing experimental data showing structural plasticity during *FSN* formation. Therefore, the aim of this paper is to bring together experiment and theoretic on definition plasticity of *FSN* network.

The goal of this paper is on defining a base for a *FSN* plasticity theory to analyze how the correlation between folksonomy tags and uri web resource can evolve as time evolve and so to have the base predict how the folksonomy tags will be chosen by the users in a future work. To perform such goal it is necessary a systematic mathematical analysis on deformation and fracture for the ontology matching that is on the base of the *FSN* [3].

On this paper is given a brief discussion on some simple problems of nonlinear behavior of the Folksodriven Ontology Matching of the *FSN* with some simple models and by extending results in the study of linear regime. Of course, these discussions are not complete, which may provide some hints for further development of the area.

The paper is arranged as follows. First, we discuss some experimental results on the nonlinear ontology matching behavior of Folksodriven tags (*FD tags*), then we describe a possible plastic constitutive equation of the ontology. Here we focus on the discussion concerning plastic flow around crack tip for some n-dimensional lattice representation of *FSN*.

In view of the difficulty for setting up the equations, we turn to introduce nonlinear elastic constitutive equations of *FD tags*.

## 2 Folksodriven Notation

$$(1) \quad FD := (C, E, R, X)$$

As stated in [1, 2] we consider a Folksodriven tag as a tuple (1) defined by finite sets composed by:

- *Formal Context* (C) is a triple *C:=(T, D, I)* where the objects *T* and the attributes *D* are sets of data, deducted with *Natural Language Processing (NLP)* techniques, and *I* is a relation between *T* and *D* [4] – see 3 (*Lattice of Folksodriven Structure Network*);
- *Time Exposition (E)* is the clickthrough rate (*CTR*) as the number of clicks on a *Resource (R)* divided by the number of times that the *Resource (R)* is displayed (impressions);
- *Resource (R)* is represented by the uri of the webpage that the user wants to correlate to a chosen tag;
- *X* is defined by the relation *X = C × E × R* in a Minkowski vector space [5] delimited by the vectors *C*, *E* and *R*.

## 3 Lattice of Folksodriven Structure Network

A *Folksodriven Structure Network (FSN)* consists of folksonomy tags (e.g. collections of tags about news websites) which are arranged regularly in Minkowsky vector space [5] defined by *Formal Context (C)*, *Time Exposition (E)* and *Resource (R)* as defined in [1].

The arrangement is a repetition of the smallest unit about a subject (e.g.: news websites regarding football match, airplane traffic, etc.), called a unit cell, resulting in the periodicity of a complete *FSN*. The frame of the periodic arrangement of centers of time exposition (*E*) of Folksodriven tags (*FD tags*) is called a lattice. Thus, the properties of corresponding points of different cells in a *FSN* are the same. The positions of these points can be defined in a Minkowsky vector space. A *FSN* is an n-dimensional array of folksonomy forming a regular lattice. The *FSN* lattice represents a network that correlate different *FD tags* and so different folksonomy tags (see *2 Folksodriven Notation* and [1]) chose by a user on a group of websites (here we consider a group of new websites). So a *FSN lattice* performs an Ontology Matching between different and unordered Folksodriven tags (*FD tags*) – that are hard to categorize.

The *FSN lattice* is designed considering its vertices and *FD tags* corresponding to the nodes of the mesh and its edges correspond to the ties between the nodes and so the *FD tags*. According to mathematic a lattice in $R^n$ is a discrete subgroup of $R^n$ which spans the real vector space $R^n$. Every lattice in $R^n$ can be generated from a basis for the vector space by forming all linear combinations with integer coefficients. A lattice may be viewed as a regular tiling of a space by a primitive cell [6].

We consider a "Bethe lattice" to construct the lattice of *FSN*, we begin with a *FD tag* at the origin and then make a bond from the origin to each of *z* neighbors *FD tag* (where *z* is called the "coordination number"). The origin is called generation 0, while the *z* new *FD tags* are called generation 1. Then, for each *FD tag x* in generation *1*, we create *z-1 FD tags* and join each of these new *FD tags* to *x* with a bond. All of these *z(z-1)* new *FD tags* form generation *2*. Then we repeat this process indefinitely. A picture of the first four generations of a Bethe lattice with *z=3* is shown in Fig. 1.

Others methods can be chosen to construct a Bethe lattice (i.e. the difference in generation number as the distance function on the Bethe lattice).

A "Bethe lattice" can be seen as a tree-like structure emanating from a central node, considered as the root or origin of the lattice, with all the nodes arranged in shells around the central one.

The number of nodes in the *k*th shell is given by:

$$(2) \quad N_k = z(z-1)^{k-1} \quad \text{for } k > 0$$

The square root of the difference between generations is taken as distance on the Bethe lattice, in which case, the second moment of the distance behaves asymptotically as it does on $Z^d$

## 4 Folksodriven Structure Network Plasticity

*FSN*s have regular elements, like normal networks. But these elements fit together in ways which never properly repeat themselves. The two-dimensional equivalent is known as Penrose tiling (Fig. 2), after Sir Roger Penrose, a British mathematician who put this form of geometry on a formal footing. Penrose created "aperiodic" tiling patterns that never repeated themselves; work that he suspects was inspired by Kepler's drawings. Penrose tiling [7] has, however, been widely used in the past for decoration.

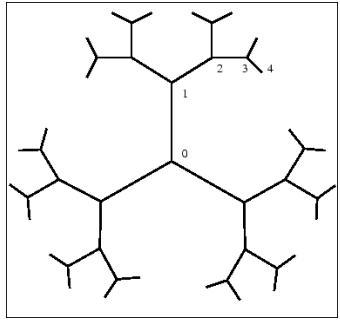 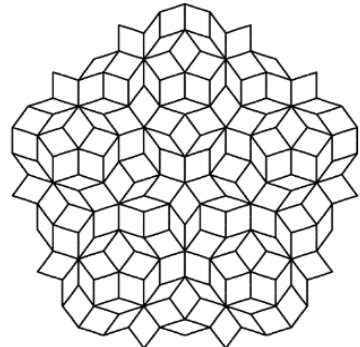

**Fig. 1.** Bethe lattice with $z=3$, the first *4* generations of *FD tags* are shown.

**Fig. 2.** Penrose tiling

A three-dimensional Penrose tile equivalent is used in chemistry for the quasicrystals [8, 9, 10] having revolutionized materials science.

While *FSN* considers n-dimensions, virtually infinite dimensions, depending on the network connection. *FSN* deals mostly with how something evolves over time. Space or distance can take the place of time in many instances. *FSN* happens only in deterministic, nonlinear [1], dynamical systems [2].

For the conventional mathematical models the nonlinear behavior means mainly plasticity. In the study on the classical plasticity there are two different theories, one is the macroscopic plasticity theory and the other is based on the mechanism of motion of dislocation, and in some extent can be seen as a "microscopic" theory.

The lack of both enough macro- and micro-data [11] makes it difficult to verify the Folksodriven plasticity. Currently the macroscopic experiments have not been properly undertaken. While there are some limited data from works on the mechanism in microscopy of the plasticity. For this reason a systematic mathematical analysis on deformation and fracture for the ontology matching is not extensible available so far.

## 5   Macroscopic behavior of plastic FSN

We observed the variations on the lattice *FSN* using parameters defined on the *FD* structure (defined in *2 Folksodriven Notation*).

---

[1] *Nonlinear* means that output isn't directly proportional to input, or that a change in one variable doesn't produce a proportional change or reaction in the related variable(s). In other words, a system's values at one time aren't proportional to the values at an earlier time.

[2] A *dynamical system* is anything that moves, changes, or evolves in time. Hence, chaos deals with what the experts like to refer to as dynamical-systems theory (the study of phenomena that vary with time) or nonlinear dynamics (the study of nonlinear movement or evolution).

We consider the plasticity as the capacity of *FSN lattice* to vary in developmental pattern, by the interaction of the *FD tags* and the web source environment, or in user behavior according to varying user choose of the *FD tags*.

The so-called *FSN plasticity theory* here developed is based on classical theory [12] considering the mechanism of motion of dislocation, and in some extent can be seen as a "microscopic" theory. The difficulty for *FSN* plasticity lies in lack of both enough macro- and micro-data.

At medium and low time exposition (*E*), Folksodriven formal context (*C*) exhibit few points of dispersion, but those present plasticity-ductility at high time exposition (*E*). Near the high stress concentration zone on formal context (*C*), e.g. around dislocation core or crack tip, plastic flow appears.

The connection between the structural defects in *FSN* and plasticity was observed in experiments where plastic deformation of a Folksodriven Structure Network (*FSN*) was induced by a discontinuity in the otherwise normal lattice structure of the network. Salient structural features are presented in the *FSN*.

Some experimental data have been obtained observing tagging on news websites, whose posses a very high time exposition (*E*).

Fig. 3 shows the stress-strain curves of the *FSN*: at strain rate of 10^(-2) time exposition (*E*) the ductile range sets on formal context (*C*) in at about 105% corresponding to a homologous time exposition (*E*) of about 0.84. At lower strain rates of 10^(-3) *E* and below limited ductility can be observed down to formal context about 72%. As another class of *FSN* the stress train curves measured by experiments are similar to those given by Fig. 4.

The intrinsic equation based on the experimental data of discontinuity density of formal context (*C*) and dislocation time exposition (*E*), e.g. the plastic strain rate and the applied stress in a power-law form [8] can be described as in (3) where *B* and *m* are time exposition dependent parameters and $s\hat{}$ is the internal variable that can be considered as a reference stress representing the current structural state of the *FSN* and is used to accommodate the model to the description of different kind of websites observed [13].

$$(3) \qquad \dot{\varepsilon} = B\left(s/\hat{s}\right)^m$$

Combining relevant information, formula (1) can be used to well predict experimental curves, e.g. recorded by Fig. 3.

The current parameters B, m and $s\hat{}$ were observed for dislocation model, which, from the angle of methodology, are different from those adopted in the classical plasticity, where they are measured from pure macroscopic approach rather [6].

Unfortunately there is lack of the comprehensive macroscopic experimental data (i.e. the data arising from multi-clustering context condition) for *FSN* plasticity so far. An interesting topic for plastic deformation studies is on periodic lattice directions of the *FSN*. Therefore the influences of periodicity and quasi-periodicity on the plastic deformation can be directly revealed on one ontology by investigating the plastic properties in different deformation geometries - an anisotropic behavior is expected.

The experimental data for stress-strain curves for *FSN* news websites are depicted in Fig. 4, it presents plastic properties depicted by distinguishing nature of *FSN lattice* between the two phases. Hence, an anisotropic behavior for the *FSN* is leaded by the influences of periodicity and quasi-periodicity on the plastic deformation.

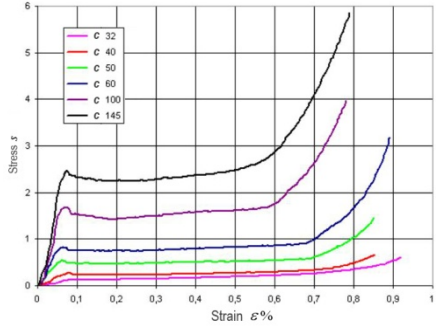 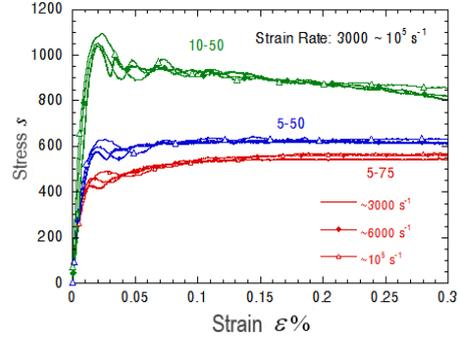

**Fig. 3.** Stress-strain curves of the *FSN* in the high formal context (*C*) range 32-145 at strain rate depicted as percentage respect the time exposition (*E*).

**Fig. 4.** Stress-strain curves of the *FSN* for 3 different ontology matching at strain rate 0-0.3 of the time exposition (*E*).

## 6 Proposed scheme of plastic constitutive equations

From the suitable experimental data the formula (3) is depicted to give support in studying plastic constitutive equations of *FSN*.

According to a macroscopically point of view, the result is obtained for unique ontology condition, as at present there is lack of data on multiple ontology condition. We assume that, if sufficient experimental data are available on "yield surface" for some *FSN*, then we can obtain an equation for the yield surface such as in (4) where $Seff = Se + f(Kij)$ denotes a generalized effective stress where: $S_e$ represents the part coming from phonon[3] stresses, $S_{ij}$ and $f(Kij)$ the part coming from phason[4] stresses $K_{ij}$.

$$(4) \quad \Omega = s_{eff} - Y = 0$$

$$(5) \quad Y = s_Y = const$$

$$(6) \quad Y = Y(k)$$

The equation (4) represents the initial yield surface if *Y* is defined as in (5) where $S_y$ is the unique surface yield limit of the *FSN*. After all if we consider *Y* related to deformation history (6) depicted by the parameter *k*, then (4) describes the evolution law of *FSN* deformation.

When the yield surface like (4) is available it is possible to construct the (7) plastic constitutive equations.

$$(7) \quad \begin{cases} \dot{\varepsilon}_{ij} = \dfrac{1}{K(s_{eff})} \dot{s}_{eff} \dfrac{\partial \Omega}{\partial s_{ij}} \\ \dot{w}_{ij} = \dfrac{1}{K(s_{eff})} \dot{s}_{eff} \dfrac{\partial \Omega}{\partial K_{ij}} \end{cases}$$

Where $\Omega$ is the above mentioned yield surface function. The variation rate of the quantities is denoted by the dot over physical quantities. $K(S_{eff})$ represents the clustering modulus of the *FSN*, which can be calibrated by a test of a simple stress-strain state, as given by (2).

The elastic-plastic constitutive equation will be obtained by adding to (7) the elastic constitutive relationship that was introduced in previous paragraphs.

A complete constitutive law of equations may be determined by (7) that represent the incremental plastic equations that describe effect of loading/unloading state within the deformation history process.

A relative simpler constitutive law relative to total plasticity theory could define the effective stress $S_{eff}$ and the effective strain $\varepsilon_{eff}$. The stress $S_{eff}$ was introduced above, while the effective strain $\varepsilon_{eff}$ has a similar definition and consists of phonon strains $\varepsilon_{ij}$ as well as phason strains $W_{ij}$. Then between the strains and stresses we can express the following relations.

$$(8) \quad \begin{cases} \varepsilon_{ij} - \dfrac{1}{3}\varepsilon_{kk}\delta_{ij} = \dfrac{3\,\varepsilon_{eff}}{2s_{eff}}\left(s_{ij} - \dfrac{1}{3}s_{kk}\delta_{ij}\right) \\ w_{ij} - \dfrac{1}{3}w_{kk}\delta_{ij} = \dfrac{3\,\varepsilon_{eff}}{2s_{eff}}\left(K_{ij} - \dfrac{1}{3}K_{kk}\delta_{ij}\right) \end{cases}$$

In which we can assume:

$$\varepsilon_{kk} = \varepsilon_{xx} + \varepsilon_{yy} + \varepsilon_{zz}$$
$$w_{kk} = w_{xx} + w_{yy} + w_{zz}$$
$$s_{kk} = s_{xx} + s_{yy} + s_{zz}$$
$$K_{kk} = K_{xx} + K_{yy} + K_{zz}$$

$$\varepsilon_{eff} = \begin{cases} \varepsilon_{eff}^{(e)} & s_{eff} < s_0 \\ A\left(s_{eff}\right)^n & s_{eff} > s_0 \end{cases}$$

We consider $S_0$, $A$ and $n$ as ontology constants of the *FSN*, which can be measured through a unique ontology test: $\varepsilon_{(e)eff}$ represents the quantity at elastic deformation stage and $S_0$ the unique ontology tensile yield stress.

The equations (8) are nonlinear elastic constitutive equations so they cannot describe deformation history, while equations (7) can do it being plastic constitutive equations related to *FSN*. Equations (7) or (8) may represent a hypothetical total plastic constitutive law or incremental plastic constitutive law for the *FSN*.

---

[3] A *phonon* is a quantum mechanical description of a special type of vibrational motion, in which a lattice uniformly oscillates at the same frequency. In classical mechanics this is known as the normal mode. The normal mode is important because any arbitrary lattice vibration can be considered as a superposition of these elementary vibrations (cfr. Fourier analysis).

[4] Similar to phonon, *phason* is associated with nodes of lattice motion, considered here as *FD tags*. However, whereas phonons are related to translation of *FD tags*, phasons are associated with FD tags rearrangements.

Actually due to the lack of enough experimental data we cannot verify if they are correct or not. We can compose the basic framework of the theory of macro-plasticity of *FSN*, in the sense of total deformation or incremental of *FSN* considering the constitutive equations (7) or (8) then by matching the equations of deformation geometry (9) and the equilibrium equations (10).

$$(9) \quad \varepsilon_{kk} = \frac{1}{2}\left(\frac{\partial u_i}{\partial x_j} + \frac{\partial u_j}{\partial x_i}\right), \quad w_{ij} = \frac{\partial w_i}{\partial x_j}$$

$$(10) \quad \frac{\partial s_{ij}}{\partial x_j} = 0, \quad \frac{\partial K_{ij}}{\partial x_j} = 0$$

Currently the equations (4), (5) and equations (7) cannot be verified due to the lack of appropriate data. Even the equations (8) have not been verified either for the same reason. It is self-evident that the possible theory is nonlinear, because the clustering network behavior is dependent with the history of deformation process and the material parameters are not constants in general.

At the moment this is the main difficult of macro-plasticity theory, cause the solution should be more difficult than that for elastic deformation.

Due to relative simplicity of the equations (8), for some simple configurations, e.g. one-dimensional ontology matching or three-dimensional ontology matching, it is possible to verify the plastic analysis by using the above proposed constitutive equations.

## 7 Computational complexity

We considered the complexity of the above procedure depending on the number of nodes in the *k*th shell (2) - which can be also expressed by the number of iterations of the algorithm. Even though it is possible to build examples where this number is large, we often observe in practice that the event where one variable leaves the active set is rare. The complexity also depends on the implementation [14].

The most widely technique used for variable selection is the Principal Component Analysis (PCA), which finds the corresponding projection by which the data show greater variability. For the square loss, the PCA remains the fastest algorithm for small and medium-scale problems on dimensionality reduction, since its complexity depends essentially on the size of the active sets. It computes the whole structure up to a certain scarcity level [15]. For mining large datasets the *FSN* could be used with the technique described in [16] – that is independent of any assumptions about the data.

## 8 Future development and conclusion

We pointed out at the beginning of this paper that in the study on the classical plasticity there are two different theories, one is the macroscopic plasticity theory, and the other is so-called crystal plasticity theory, in some extent the latter can be seen as a "microscopic" theory, which is based on the mechanism of motion of dislocation.

This paper discussed the deformation exhibiting nonlinear behavior on *FSN* based on folksonomy tags chosen by different user on web site resources, this is a topic which has not been well studied so far. The discussion in this paper shows that the nonlinear elastic constitutive equation possesses some leaning for the investigation due to lack of plastic constitutive equation at present. A constitutive law on *FSN* should be investigated towards a systematic mathematical analysis on deformation and fracture for the ontology matching.

**Massimiliano Dal Mas** is an engineer at the Web Services division of the Telecom Italia Group, Italy. His interests include: user interfaces and visualization for information retrieval, automated Web interface evaluation and text analysis, empirical computational linguistics, text data mining, knowledge engineering and artificial intelligence. He received BA, MS degrees in Computer Science Engineering from the Politecnico di Milano, Italy. He won the thirteenth edition 2008 of the CEI Award for the best degree thesis with a dissertation on "Semantic technologies for industrial purposes" (Supervisor Prof. M. Colombetti).